\documentclass[graybox,natbib]{svmult}

\usepackage{natbib}
\usepackage{makeidx}         
\usepackage{graphicx}        
\usepackage{multicol}        
\usepackage[bottom]{footmisc}

\makeindex             

\newcommand	\gtsim	{\lower.5ex\hbox{$\buildrel > \over \sim$}}
\newcommand     \ltsim  {\lower.5ex\hbox{$\buildrel < \over \sim$}}
\newcommand	\simgt	{\lower.5ex\hbox{$\buildrel > \over \sim$}}
\newcommand     \simlt  {\lower.5ex\hbox{$\buildrel < \over \sim$}}
\def    \beq       {\begin{equation}}
\def    \eeq       {\end{equation}}

\def    \Angstrom  {\,{\rm\AA}}

\def    \cm     {\,{\rm cm}}
\def    \AU     {\,{\rm AU}}

\def    \erg    {\,{\rm erg}}
\def	\g	{\,{\rm g}}

\def    \H      {{\rm H}}

\def    \K      {\,{\rm K}}

\def    \kpc    {\,{\rm kpc}}
\def    \magni  {\,{\rm mag}}
\def    \nH     {n_{\rm H}}
\def    \NH     {N_{\rm H}}
\def    \s      {\,{\rm s}}

\def	\mum	{\,\mu{\rm m}}

\def	\simali	{\sim\,}

\def	\nm     {\,{\rm nm}}
\def	\GHz    {\,{\rm GHz}}
\def	\eV     {\,{\rm eV}}
\def    \etapl  {\eta_{\rm PL}}

\def	\chiMMP	{\chi_{\rm MMP}}

\begin{document}

\title*{Nanodust in the Interstellar Medium in Comparison to the Solar System}
\author{Aigen Li \and Ingrid Mann}
\institute
{
Department of Physics and Astronomy,
University of Missouri, Columbia, MO 65211, USA;
\texttt{LiA@missouri.edu}
\and Belgian Institute for Space Aeronomy, B-1180 Brussels, Belgium;
\texttt{ingrid.mann@aeronomie.be}
}
%
%
\maketitle

\abstract{%
Nanodust, which undergoes stochastic heating 
by single starlight photons in the interstellar medium, 
ranges from angstrom-sized large molecules containing
tens to thousands of atoms 
(e.g. polycyclic aromatic hydrocarbon molecules)
to grains of a couple tens of nanometers.
The presence of nanograins in astrophysical environments
has been revealed by a variety of interstellar phenomena:
the optical luminescence, the near- and mid-infrared emission, 
the Galactic foreground microwave emission,
and the ultraviolet extinction which are ubiquitously
seen in the interstellar medium of the Milky Way and beyond.
Nanograins (e.g. nanodiamonds) have also been identified 
as presolar in primitive meteorites based 
on their isotopically anomalous composition.
Considering the very processes that lead to the detection of nanodust
in the ISM for the nanodust in the solar system shows that the observation
of solar system nanodust by these processes is less likely. 
}

\section{Introduction: The Interstellar Medium and Nanodust}
\label{sec:intro}
%
The stars in our Galaxy,  the Milky Way, are far apart
(e.g. the nearest star, Proxima Centauri, 
is at a distance of $\simali$$2.67\times 10^5\AU$ from the Sun;
1\,AU\,$\approx$\,$1.496\times 10^{13}\cm$).
The space between stars contains gaseous ions, atoms, 
molecules and solid dust grains (i.e. the interstellar medium, ISM). 
With a mean number density of 
$\simali$1\,H-atom/${\rm cm}^{3}$ it is more empty 
than the best vacuum (which has a density of 
$\simali$$10^{3}$\,molecules/${\rm cm^3}$)
that can be created on Earth.
The gas and dust of the ISM
makes up $\simali$10\%
of the total mass of the visible matter 
in the present-day Milky Way.\footnote{%
  This mass fraction is much higher for galaxies 
  at early times (since the ISM is gradually consumed 
  by star formation as galaxies evolve) 
  and much lower ($\simali$0.1\%) for elliptical galaxies. 
  }
The bulk of the heavy elements,
  including most of the interstellar silicon, magnesium, iron 
  and a large fraction of the interstellar carbon,
 are depleted from the gas phase and form submicron-sized grains,
 which make up $\simali$1\% of 
the total mass of the ISM.
In a spiral galaxy like the Milky Way, 
most of the interstellar dust and gas are concentrated
in its spiral arms and a relatively thin gaseous disk
of a thickness of a few hundred parsecs
(pc; 1\,pc\,$\approx$\,$3.086\times 10^{18}\cm$).

The ISM plays a crucial role in galaxy evolution: 
New stars form out of dusty molecular clouds 
which present a dense phase of the ISM while 
stars in late stage of evolution return gas and newly formed dust 
to the ISM (either through stellar winds 
or supernova explosions). 
The astrophysics of the ISM, from the thermodynamics
and chemistry of the gas to the dynamics of star formation
is strongly influenced by the presence of the dust.

The existence of small solid dust grains 
in interstellar space was established 
in 1930 when Trumpler showed that the stars in distant open clusters 
appear fainter than could be accounted for 
just by the inverse square law, 
and many stars in the galactic plane 
appear redder than expected from their spectral types;
he interpreted these observations in terms of
interstellar extinction and selective absorption (i.e. reddening)
caused by ``fine cosmic dust particles of various sizes''
(Trumpler 1930).
In 1956 John Platt suggested that very 
small grains or large molecules of less than 1$\nm$ in 
radius could grow in the ISM by random accretion from 
interstellar gas and could be responsible for 
the interstellar extinction. 
Today the presence of nanodust in the ISM is generally
accepted, but there are still uncertainties in interpreting
observations. 
The discussion in this chapter will focus
on the ultrasmall, nano-sized
interstellar grains. 
By nano-sized grains, or nanodust we mean grains
with a spherical radius of $a$\,$\simlt$\,10--20$\nm$ 
which undergo stochastic heating in the Galactic ISM 
(Draine \& Li 2001, Li 2004).

For a long time most of our knowledge about interstellar 
dust was derived from interstellar extinction and reddening,
and to a lesser degree from interstellar polarization
(which is caused by preferential extinction of one linear 
polarization over another by aligned nonspherical dust)
and we will discuss this in the following section (\ref{sec:extinct}).
Infrared observations from satellites started in the 1980s 
and allowed for observing the emission from
interstellar dust.
The observations indicate the presence of stochastically
heated nanodust and emitting polycyclic aromatic hydrocarbon molecules
(PAHs) discussed in
section 
(\ref{sec:IRemission}).
Other observational results are also explained with the
presence of nanodust:
the microwave emission of 
rotationally excited nanodust 
(section \ref{sec:microwave}), 
the photoluminescence of nanodust 
(section \ref{sec:ERE}), and indirect evidence
comes from the photoelectric 
heating of interstellar gas 
(section \ref{sec:photoelectric}).  
Another population of nanodust
are presolar nanograins that are identified in 
primitive meteorites and interplanetary dust and 
were present in the ISM at the time for the 
formation of the solar system
(section \ref{sec:presolargrain}).
Based on the different processes that provide evidence 
for the existences of nanodust in the ISM we then
make a comparison to the nanodust in the solar system
(section \ref{sec:SolarSystem}) 
and end with a conclusion. 
The interstellar extinction and the stochastic heating process
are elaborated in the appendix.

\section{The Interstellar Extinction}
\label{sec:extinct}

For the ISM in the solar neighbourhood 
(i.e. a few kiloparsecs from the Sun and within $\simali$100\,pc
of the galactic plane), the mean visual extinction 
per unit path-length of 
$\langle A_V/L\rangle \approx 1.8\magni\kpc^{-1}$
has long been determined quite accurately
(e.g. see Kapteyn 1904).
The wavelength-dependence of extinction
-- the interstellar extinction curve 
increases from red to blue, so that the light reaching us from 
  the stars is ``reddened'' owing to greater 
attenuation of the blue light. 
The Galactic interstellar extinction curves have now been 
measured for various sightlines over a wide wavelength
range (0.1$\mum$ $\le$ $\lambda$ $\le$20$\mum$).

\begin{figure}
\centering
\includegraphics[height=8cm,width=8cm]{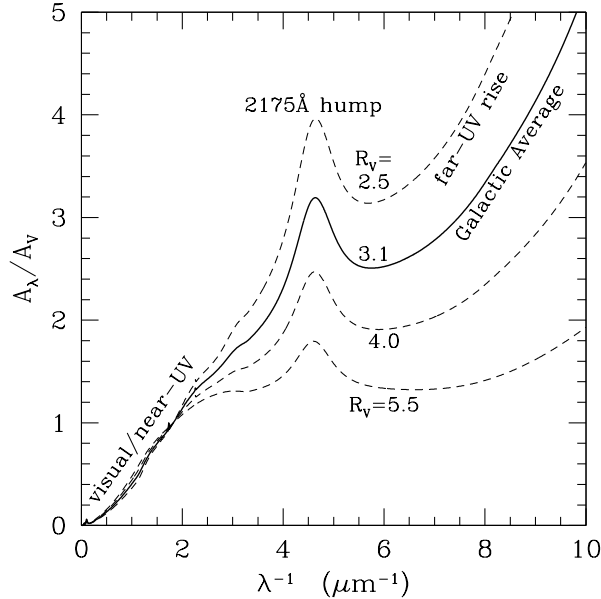}
\vspace{-2mm}
\caption{
         \label{fig:extcurv}       
        Interstellar extinction curves of
        the Milky Way ($R_V$\,=\,2.5, 3.1, 4.0, 5.5).
        There exist considerable regional variations 
        in the Galactic optical/UV extinction curves,
        as characterized by the total-to-selective 
        extinction ratio $R_V$, indicating that
        dust grains on different sightlines have 
        different size distributions. 
        }
\end{figure}

Although the extinction curves vary in shape from
one line of sight to another, they do exhibit some
common appearance (see Fig. \ref{fig:extcurv}).
The extinction curves are plotted as $A_\lambda/A_V$,
where  $A_\lambda$ is the extinction at wavelength $ \lambda$ measured in
astronomical magnitudes
and typically given relative to the extinction  $A_V$ in the visible
wavelength band.  
The extinction curves shown vs. inverse wavelength $\lambda^{-1}$
rise almost linearly from the near-infrared (IR) 
to the near-ultraviolet (UV), 
with a broad absorption bump at about 
$\lambda^{-1}$$\approx$4.6$\mum^{-1}$ 
($\lambda$$\approx$2175$\Angstrom$) 
and followed by a steep rise into the far-UV
at $\lambda^{-1}$$\approx$10$\mum^{-1}$,
the shortest wavelength at which the extinction
has been measured.  

In the wavelength range of 
0.125$\mum$\,$\ltsim$$\lambda$$\simlt$\,3.5$\mum$,
the Galactic extinction curves can be approximated by 
an analytical formula involving 
only one free parameter:
the total-to-selective extinction ratio, $R_V$
(Cardelli et al.\ 1989, see appendix).
The sightlines through diffuse gas in the Milky Way 
have $R_V\approx 3.1$ as an average value,
but there are considerable regional variations and also
the strength and width of the 2175$\Angstrom$ 
extinction bump vary markedly in the ISM 
(see Xiang et al.\ (2011) and references therein).
Lower-density regions have 
a smaller $R_V$, 
a stronger 2175$\Angstrom$ bump 
and a steeper far-UV rise 
at $\lambda^{-1}$\,$>$\,4$\mum^{-1}$,
while denser regions have a larger $R_V$, 
a weaker 2175$\Angstrom$ bump 
and a flatter far-UV rise.

The exact nature of the carrier of this bump remains unknown 
since its first discovery
nearly half a century ago (Stecher 1965). 
It has been postulated to be nano carbon particles 
(e.g. Duley \& Seahra 1998) 
or PAHs (Joblin et al.\ 1992, Li \& Draine 2001a, 
Cecchi-Pestellini et al.\ 2008, Steglich et al.\ 2010).

The far-UV part ($\lambda$\,$\simgt$\,6$\mum^{-1}$)
of the Galactic interstellar extinction 
continues to rise up with shorter wavelength to $\lambda$\,=\,0.1$\mum$ 
and there does not appear to be any evidence 
of saturation even at this wavelength.\footnote{%
  However, the Kramers-Kronig dispersion relation requires 
  that the far-UV extinction rise with inverse wavelengths 
  must turn over at some smaller wavelengths 
  as the wavelength-integrated extinction 
  must be a finite number (see Purcell 1969).  
  }
Since it is generally true that a grain absorbs 
and scatters light most effectively at wavelengths 
comparable to its size $\lambda \approx 2\pi a$,
we can therefore conclude that there must be 
appreciable numbers of ultrasmall grains with 
$a$\,$\simlt$\,$0.1\mum/2\pi$\,$\approx$\,$16\nm$.
In the far-UV wavelength
range, the grains of a couple of nanometers are in the Rayleigh regime
(i.e. $2\pi a/\lambda$$\ll$1) and their extinction cross sections 
per unit volume, $C_{\rm ext}(a,\lambda)/V$,
are independent of size.
Hence the far-UV extinction indicates the presence of nanodust, but 
does not allow to constrain the sizes 
of the nanodust in the ISM.




\section{Emission Brightness from Nanodust and Molecules} 
         \label{sec:IRemission}

While the far-UV interstellar extinction requires the
presence of a population of nano-sized grains 
in the ISM, the near- and mid-IR\footnote{%
  We here use ``near-IR'' for wavelengths
  1$\mum$\,$\ltsim$$\lambda$$\simlt$\,12$\mum$,
  ``mid-IR'' for 12$\mum$\,$\ltsim$$\lambda$$\simlt$\,60$\mum$,
  and ``far-IR'' for 60$\mum$\,$\ltsim$$\lambda$$\simlt$\,1000$\mum$.
  }  
emissions provide constraints 
of their size and composition.
The nanodust is seen by its stochastic heating and by
its characteristic emission. 

A dust particle in space 
is subject to substantial 
temporal fluctuations in temperature, if
(i) its heat content is smaller than or comparable 
    to the energy of a single stellar photon 
    (Greenberg 1968), and
(ii) the photon absorption rate is smaller than
the radiative cooling rate (Li 2004).
In the diffuse ISM, nanodust is 
stochastically heated by single photons 
to temperatures much higher than 
its ``equilibrium'' temperature 
(even though an ``equilibrium'' temperature 
is not physical for a stochastically-heated nanograin, 
it can still be {\it mathematically} determined from 
the energy balance between absorption and emission).

\begin{figure}
\centering
\includegraphics[angle=270,width=8.8cm]{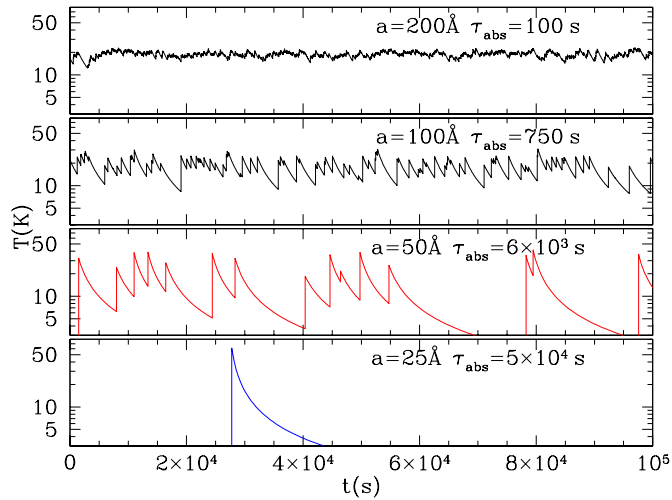}
\vspace{-2mm}
\caption{
         \label{fig:T_vs_t} 
         The time evolution of grain temperature within a day 
         ($\simali$8.6$\times$10$^4\s$)
         for four PAH/graphitic grains exposed to the solar 
         neighbourhood interstellar radiation field.
         $\tau_{\rm abs}$ is the mean time between photon absorptions.
         Grains with radii $a$\,$\gtsim$20\,nm
         obtain an equilibrium temperature  
         (i.e. their temperature does not fluctuate much with time)
         due to their large photon absorption rates $1/\tau_{\rm abs}$
         (because of their large absorption cross sections)
         and their large heat capacities 
         (so that a single photon cannot significantly
          raise their temperature). 
         For grains with radii $a$\,$\ltsim$5\,nm
         their temperature strongly fluctuates with time
         due to their small heat capacities
         (so that a single photon can result in 
          an appreciable raise of their temperature)
         and their small photon absorption rates
         (so that they could have already cooled down 
          before the absorption of another photon).
         Taken from Draine (2003).
         }
\vspace{-2mm}
\end{figure}

Figure \ref{fig:T_vs_t} illustrates the time evolution of
grain temperature within a day for four PAH/graphitic grains
exposed to the solar neighbourhood interstellar radiation field
(Draine 2003). We see that for grain radii $a$\,$\gtsim$\,20$\nm$, 
individual photon absorptions are relatively frequent, 
and the grain heat capacity is large enough that
the temperature excursions following individual photon absorptions
are relatively small; it is reasonable to approximate 
the grain temperature as being constant in time.
Grains with radii, $a\ltsim5$\,nm, however, 
cool appreciably in the time between photon absorptions; 
as a result, individual photon absorption events raise 
the grain temperature to well above the mean value.
A PAH molecule of 
100 carbon atoms (corresponding to a size of 
$\simali$6$\Angstrom$)\footnote{%
  The term ``PAH size'' refers to the radius $a$ of 
  a spherical grain with the same carbon density as 
  graphite ($2.24\g\cm^{-3}$) and containing
  the same number of carbon atoms $N_{\rm C}$: 
  $a$\,$\equiv$\,1.286\,$N_{\rm C}^{1/3}\Angstrom$. 
  }
will even be heated to $T$\,$\approx$\,785$\K$ by a photon of
$h\nu$\,=\,6$\eV$, while its ``equilibrium'' temperature 
would just be $\simali$22$\K$. The stochastic heating is
further discussed in the appendix.

\begin{figure}
\centering
\vspace{-6mm}
\includegraphics[width=9cm]{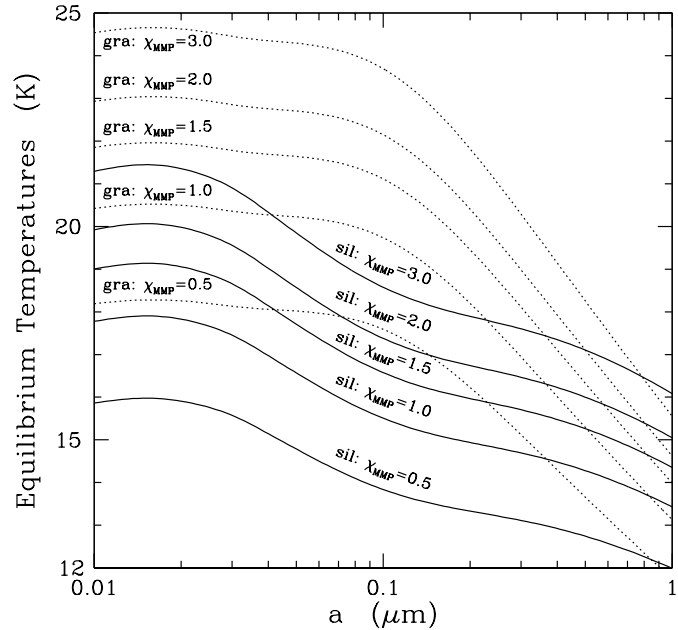}
\vspace{-2mm}
\caption{
         \label{fig:Tss}
	Equilibrium temperatures for graphite (dotted lines) 
        and silicate grains (solid lines) in environments with
        various starlight intensities ($\chiMMP$, in units of 
        the Mathis, Mezger, \& Panagia (1983)
        solar neighbourhood interstellar radiation field). 
        Both silicate and graphitic grains larger than 10--20$\nm$ 
        attain equilibrium temperatures
        in the range of 12$\K$$<$\,$T$\,$<$\,25$\K$
        and these grains are too cold to emit appreciably 
        at $\lambda$\,$<$\,60$\mum$.
        The equilibrium temperatures of grains with $a$\,$<$\,20$\nm$
        do not depend on their size.
        Taken from Li \& Draine (2001a).    
        }
\end{figure}

Andriesse \& de Vries (1976)
presented the first IR emission evidence 
for nanodust in the dust cloud 
in M\,17, a star-forming nebula. 
They found that the 8--20$\mum$ emission spectrum is similar 
over a distance of $\simali$2$^{\prime}$ through the source,
suggesting a constant dust temperature.   
Since large, submicron-sized grains  
would attain equilibrium temperatures that decrease 
with distance from the illuminating source, 
Andriesse (1978) interpreted this as due to 
stochastically heated grains of $\simali$1\,nm.

A more definite piece of observational evidence was provided
by Sellgren et al.\ (1983). 
In near-IR observations 
of three visual reflection nebulae (i.e. NGC 7023, 2023, and 2068), 
they discovered that each nebula has extended near-IR emission 
consisting of emission features at 3.3 and 3.4$\mum$ 
and a smooth continuum characterized by a color temperature 
$\simali$1000$\K$. Both the 3.3$\mum$ feature
and the color temperature of the continuum 
show very little variation from source to source 
and within a given source with distance 
from the central star. 
Sellgren (1984) argued that 
this emission could not be explained by thermal emission 
from dust in radiative equilibrium with the central star 
since otherwise the color temperature of this emission 
should fall off rapidly with distance from the illuminating 
star; instead, she proposed that this emission is due
to stochastically heated nanodust. 


\begin{figure}
\centering
\includegraphics[angle=270,width=10.8cm]{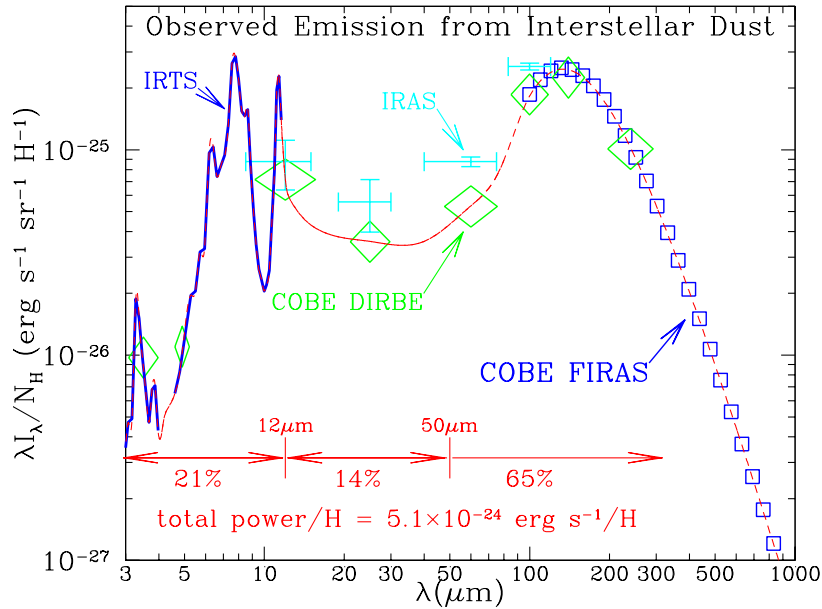}
\vspace{-2mm}
\caption{
         \label{fig:draine_irem}
        Observed diffused emission of interstellar dust from 
        the {\it Infrared Astronomical Satellite} (IRAS),
        the {\it Infrared Telescope in Space} (IRTS),
        the {\it Spitzer Space Telescope} (Spitzer),
        and the {\it Cosmic Background Explorer} (COBE) satellite 
        with its {\it Far Infrared Absolute Spectrophotometer} (FIRAS)
        and its {\it Diffuse Infrared Background Experiment} (DIRBE). 
    The total IR emission normalized to the hydrogen column density $\NH$,
         is $\simali$$5.1\times 10^{-24}\erg\s^{-1}\H^{-1}$. 
         The crosses denote observations from 
         IRAS (Boulanger \& Perault 1988),
         squares from COBE-FIRAS (Finkbeiner et al.\ 1999), 
         diamonds from COBE-DIRBE (Arendt et al.\ 1998), 
         and the heavy curve from IRTS 
         (Onaka et al.\ 1996, Tanaka et al.\ 1996).
         The ``UIR'' bands at 3.3--11.3$\mum$
        (generally attributed to PAHs) 
        account for $\simali$20\% of the total IR emission,
        the emission from nanodust at $\lambda\simlt60\mum$        
        accounts for $\simgt$\,35\% and the 
        submicron-sized grain population emits mostly
        at $\lambda\simgt60\mum$ and accounts for $\simlt$\,65\%.
         Taken from Draine (2003). 
        }
\end{figure}

Infrared observations from satellites are made since the 1980s and
today the ISM is known for a broad spectral interval
(Figure \ref{fig:draine_irem}).
Initially the diffuse ISM 
emission observed at 12$\mum$ and 25$\mum$ 
suggested
the presence of nanodust,
since it exceeds the emission 
from large grains with $\simali$12--25$\K$ 
thermal equilibrium temperature (Fig. \ref{fig:Tss})
by several orders of magnitude
(Boulanger \& Perault 1988).
%
Later observations showed broadband 
emission at 3.5$\mum$ and 4.9$\mum$ (Arendt et al.\ 1998),
as well as  at 3.3, 6.2, 7.7, 
8.6, and 11.3$\mum$
(Onaka et al.\ 1996, Tanaka et al.\ 1996, Mattila et al.\ 1996).
These emission features are collectively referred to 
as the ``UIR'' bands, and often attributed to PAH molecules
(L\'eger \& Puget 1984, Allamandola et al.\ 1985).
%
%


\begin{figure}
\centering
\includegraphics[angle=270,width=11.2cm]{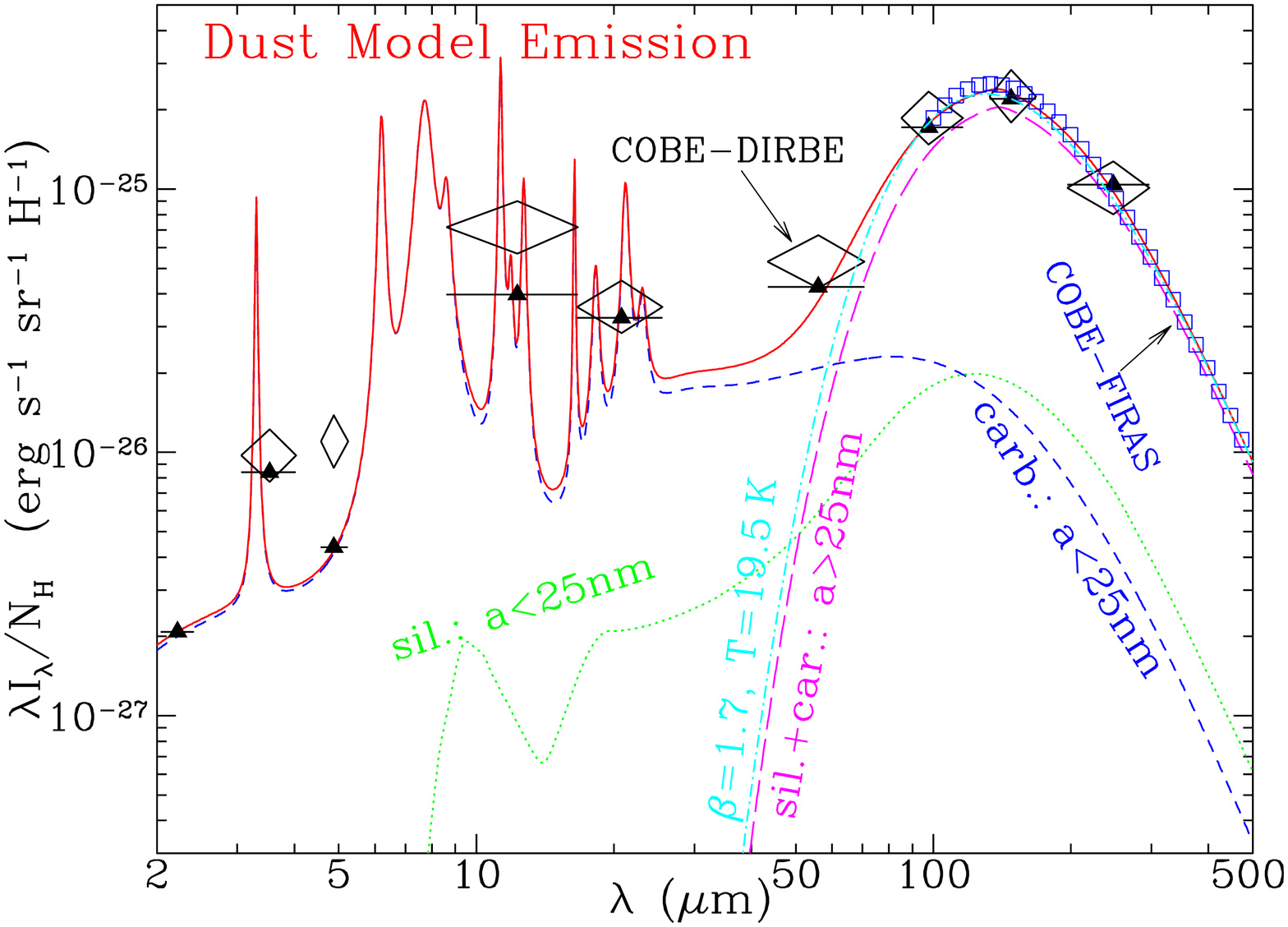}
\vspace{-2mm}
\caption{
         \label{fig:mod_irem} 
        Comparison of the ISM dust model to the observed
        emission from the diffuse ISM. 
        Dotted green line: small silicate grains with $a$\,$<$\,25$\nm$;
        Dashed blue line: small carbonaceous grains 
        (graphite and PAHs) with  $a$\,$<$\,25$\nm$;
        Long dashed magenta line: the sum of big silicate
        and graphite grains with  $a$\,$>$\,25$\nm$;
        Dot-dashed cyan line: a modified gray body of
        $I_\lambda \propto \lambda^{-\beta} B_\lambda(T)$
        with $\beta\approx 1.7$ and $T\approx 19.5\K$
        approximating the far-IR emission at $\lambda$\,$>$\,100$\mum$
        determined by DIRBE-FIRAS. 
        Triangles show the model spectrum (solid curve)
	convolved with the DIRBE filters.
        Observational data are from 
        DIRBE (diamonds; Arendt et al.\ 1998), 
	and FIRAS (squares; Finkbeiner et al.\ 1999).
        For abbreviations see caption to Fig.  \ref{fig:draine_irem}.
        Taken from Li \& Draine (2001a).
        }
\end{figure}

We now discuss a particular model to explain the different ISM dust
observations. 
Figure \ref{fig:mod_irem} shows a comparison of 
the observed emission with the emission calculated 
from the silicate-graphite-PAH model (Li \& Draine 2001a).
In this model, $\simali$15\% of the carbon is locked up
in the PAH component and the near-IR 
and 
UIR spectrum are best reproduced by PAHs with
a log-normal size distribution peaking at $a$\,$\simali$6$\Angstrom$, 
corresponding to $\simali$100 carbon atoms.
At $\lambda$\,$\simlt$\,60$\mum$ the emission is predominantly
from small carbonaceous grains (graphite and PAHs)
with $a$\,$<$\,25$\nm$. 
Even at $\lambda$\,=\,60$\mum$, the grains with $a$\,$<$\,25$\nm$  
contribute $\simali$70\% of the total emission power of 
the diffuse ISM.
The large silicate and graphite grains of $a$\,$>$\,25$\nm$
together dominate the emission at $\lambda$\,$>$\,60$\mum$,
accounting for $\simali$35\% of the total IR power.
The far-IR emission at $\lambda$\,$>$\,100$\mum$
can be closely approximated by a modified 
black body of $I_\lambda \propto \lambda^{-\beta} B_\lambda(T)$
with $\beta\approx 1.7$ and $T\approx 19.5\K$
(see Draine 1999) or by model emission calculated from 
large grains with $a$\,$>$\,25$\nm$. 
The emission at 
$\lambda$\,$=$\,12, 25$\mum$ and shorter wavelengths
(as well as part of the 60$\mum$ band) 
cannot be explained by large grains 
and it requests a population of stochastically heated
nanodust.


\vspace{-3.0mm}
\section{Microwave Emission from Rotationally-Excited Nanodust
            \label{sec:microwave}}
\begin{figure}
\centering
\vspace{-2mm}
\includegraphics[angle=270,width=11.2cm]{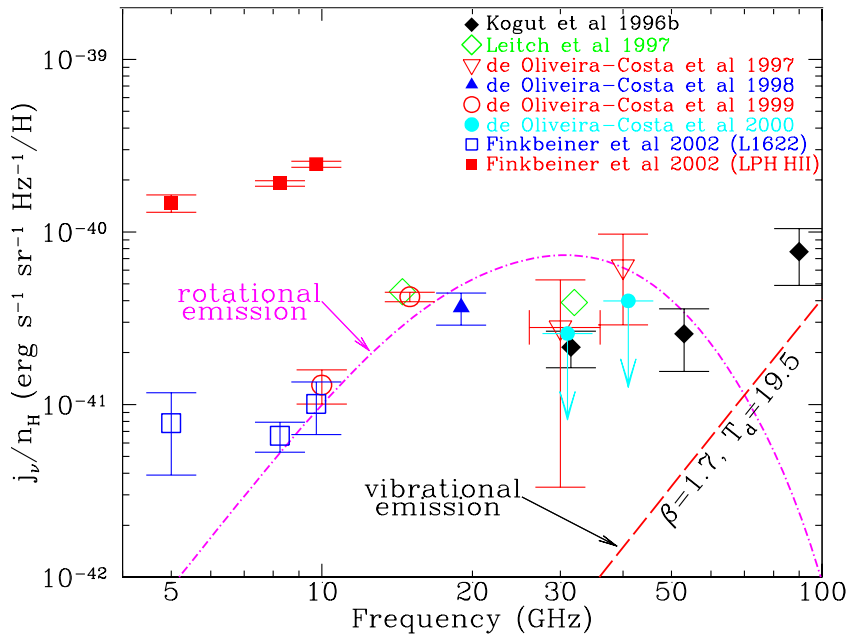}
\vspace{-2mm}
\caption{
         \label{fig:btd_spinningdust}
         Comparison of the ``anomalous'' Galactic foreground 
         microwave emission in the 10--100$\GHz$ frequency region
         with the rotational electric dipole emission 
         calculated from fast-spinning nanograins (i.e. PAHs) 
         which account for the ``UIR'' vibrational emission
         (see Figures \ref{fig:draine_irem}, \ref{fig:mod_irem}).
	 Symbols: observational determinations of 
         ``anomalous microwave emission''.
         Dot-dashed line: model rotational emission spectrum
                          of nanodust (Draine \& Lazarian 1998).
         Dashed line: low-frequency tail of the 
                      emission from large grains 
                      (mostly with $a$\,$>$\,25$\nm$) 
         (also see Figures \ref{fig:draine_irem},\,\ref{fig:mod_irem}).
         Taken from Draine (2003).
         }
\end{figure}

\vspace{-2.0mm}
Numerous sensitive experiments to map the microwave sky 
have revealed unexpected emission at 10--100$\GHz$ frequencies
(see Figure \ref{fig:btd_spinningdust}).
The spectral variation and absolute value of this  ``anomalous'' 
component of the diffuse foreground microwave emission 
%
are very different from those of 
the traditional diffuse emissions at these frequency ranges
(e.g. the free-free, synchrotron, and thermal dust emission 
have power-law--like spectra at microwave frequencies)
and can not easily be explained with them.

The spatial distribution of this microwave emission 
is correlated with interstellar dust emission
at 100$\mum$ and 140$\mum$ 
(see Draine 2003), and even better 
correlated 
with the mid-IR emission
(Casassus et al.\ 2006, Ysard et al.\ 2010, Vidal et al.\ 2011).

This suggests its origins from the dust, but extrapolating 
the 100--3000$\mum$ far-IR emission 
of large dust ($a$\,$>$\,25\,nm) 
to the microwave frequencies 
provides values far below the observed microwave emission 
(see Figure \ref{fig:btd_spinningdust}).
For these reasons, the electric dipole radiation from 
rapidly spinning nanograins has become the best explanation 
for the ``anomalous'' microwave emission.

A spinning grain with an electric dipole moment {\bf $\mu$} 
radiates power $P =2\omega^4\mu^2 \sin^2\theta/3c^3$,
where $\theta$ is the angle between the angular velocity $\omega$ 
and ${\bf \mu}$, and $c$ is the speed of light. 
The angular velocity $\omega$\,=\,$J/I$, 
where $J$ is the grain angular moment, 
and $I$ is the moment of inertia of the grain.
For spherical grains, $I\propto a^5$. 
Hence angular velocity steeply decreases with grain size and
in interstellar environments
only nanograins can be driven to
rotate fast enough to emit at microwave frequencies.   
For a PAH grain of radius $a$\,=\,1\,nm in the diffuse ISM,
$J$ peaks at $\simali$2000\,$\hbar$ (see Draine \& Lazarian 1998).

As described by Draine and his coworkers 
(Draine \& Lazarian 1998, Hoang et al.\ 2010),
a number of physical processes, 
including collisions with neutral atoms and ions, 
``plasma drag'' (due to interaction of the electric 
dipole moment of the grain with the electric field 
produced by passing ions), and absorption and emission 
of photons, can drive nanograins to rapidly rotate, 
with rotation rates reaching tens of GHz. 
The rotational electric dipole emission 
from these spinning nanograins, 
the very same grain component (i.e. PAHs)
required to account for the ``UIR'' emission 
and the observed
12 and 25$\mum$ continuum emission, 
was shown to be capable of accounting for 
the ``anomalous'' microwave emission
(Draine \& Lazarian 1998; 
see Figure \ref{fig:btd_spinningdust}). 
  Vidal et al.\ (2011) found that the microwave emission
  at 31\,GHz of the LDN\,1780 translucent cloud
  correlates better with the 12, 25$\mum$ emission
  than with the 100$\mum$ emission, which supports
  that emission at this frequency originates from nanodust.

We should note that although the electric dipole radiation
from spinning nanodust provides the best explanation for 
the ``anomalous'' microwave emission, other physical mechanisms 
(e.g. hot free-free emission, hard synchrotron radiation, 
or magnetic dipole emission)
could still be contributing at some level
(e.g. see Draine \& Lazarian 1999).


\vspace{-3.0mm}
\section{Extended Red Emission: Photoluminescence 
            from Nanodust \label{sec:ERE}}
\begin{figure}
\centering
\vspace{-4mm}
\includegraphics[width=9.0cm]{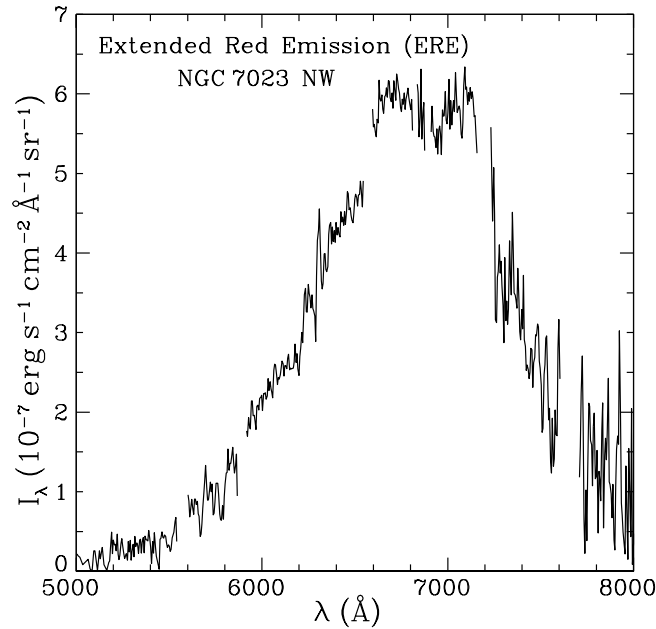}
\vspace{-4mm}
\caption{
         \label{fig:ERE_NGC7023}
        Observed photoluminescence spectrum 
        -- the ``Extended Red Emission'' (ERE)
        arising from an unidentified nanodust species 
        in the north-west (NW) filament
        of NGC 7023, a reflection nebula.
        Data taken from Witt \& Vijh (2004).
        }
\end{figure}

\vspace{-2.0mm}
%
Dust {\it emission} at optical wavelengths,
not expected from its vibrational excitation, 
is also seen in the red part of the visible spectra 
of a wide variety of dusty environments 
(which cannot be accounted for just by dust scattering alone). 
This emission is termed ``extended red emission'' (ERE). 
It is characterized by a broad, featureless band between 
$\simali$5400$\Angstrom$ and 9500$\Angstrom$, with a width 
of $600\Angstrom \ltsim {\rm FWHM} \ltsim 1000\Angstrom$
and a peak of maximum emission at
$6100\Angstrom \ltsim \lambda_{\rm p} \ltsim 8200\Angstrom$, 
depending on the physical conditions of the environment 
where the ERE is produced 
(see Figure \ref{fig:ERE_NGC7023}). 
   The peak wavelength $\lambda_{\rm p}$ 
   varies from source to source and within
   a given source with distance from the illuminating star.
   The ERE width appears to increase as $\lambda_{\rm p}$
   shifts to longer wavelengths (Darbon et al.\ 1999).
   This suggests that the ERE carriers can be easily 
   modified (or destroyed) by intense UV radiation,
   as $\lambda_{\rm p}$ is related to the size of
   the carrier grain.

The ERE has been seen in various astrophysical regions:
the diffuse ISM of our Galaxy and other galaxies, 
reflection nebulae, planetary nebulae, and HII regions,
which, in terms of UV photon densities, span a range of 
six orders of magnitudes, and in terms of dust, represent
both heavily processed interstellar dust and relatively
``fresh'' dust produced in local outflows 
(e.g. planetary nebulae and the proto-planetary nebula Red Rectangle;
see Witt \& Vijh 2004).

The ERE is commonly attributed to photoluminescence (PL)
by some component of interstellar dust,
a process in which absorptions of UV photons
are followed by electronic transitions 
associated with the emission of optical 
or near-IR photons.
%
The ERE is powered by UV/visible photons,
as demonstrated by Smith \& Witt (2002)
who found that the maximum ERE intensity 
in any given environment is closely correlated 
with the density of UV photons.

The true nature of the ERE carriers still remains unknown,
although over a dozen candidates have been proposed 
over the past decades. For a proposed candidate to be valid,
it must luminesce in the visible with its spectrum matching 
that of the observed ERE. But this is not sufficient.
As many candidate materials luminesce
in the visible after excitation by UV photons, 
along with the carrier abundance, 
the efficiency for photoluminescence  -- the quantum efficiency for 
the conversion of UV photons absorbed by the ERE 
carrier to ERE photons --
represents one of the strongest constraints.

Gordon et al.\ (1998)
placed a lower limit on
the photon conversion efficiency $\etapl$ 
(measured by the number ratio of luminesced photons 
to exciting UV photons) to be approximately $(10\pm3)\%$
(also see Szomoru \& Guhathakurta 1998).  
This lower limit was derived from the correlation 
of ERE intensity with HI column density 
at high Galactic latitudes (Gordon et al.\ 1998),
with the assumption that {\it all UV absorption 
is due to the ERE carrier candidate}
(in other words, assuming that {\it all the UV photons 
absorbed by dust lead to the production of ERE}).
As there are other known absorbing interstellar 
dust components not likely associated with ERE,
the actual  luminescing efficiency must be substantially larger than 10\%,
perhaps in the vicinity of 50\% or even higher
(Smith \& Witt 2002). This poses a serious challenge
to materials once thought to be promising ERE candidates,
as their  luminescing efficiencies are $<$\,1\%
(see Wada et al.\ 2009, Godard \& Dartois 2010).

All these suggest that the ERE carriers are very likely
      in the nanometer size range because (a) in general,
      nanograins are expected to luminesce efficiently 
      through the recombination of the electron-hole pair
      created upon absorption of an energetic photon,
      since in such small systems the excited electron 
      is spatially confined and the radiationless transitions 
      that are facilitated by Auger- and defect-related 
      recombination are reduced;
and (b) small nanograins may be photolytically 
      more unstable and/or more readily photoionized in 
      regions where the radiation intensity exceeds certain 
      levels of intensity and hardness, and thus resulting in 
      both a decrease in the ERE intensity and a redshift of 
      the ERE peak wavelength.\footnote{%
             This is because 
             (i) photoionization would quench 
                 the luminescence of nanograins, and 
             (ii) the smaller grains would be selectively 
                  removed due to size-dependent 
                  photofragmentation (Smith \& Witt 2002).
          Due to the quantum confinement effect, the band gap of
          a semiconductor-like nanograin is {\it smaller} 
          (and therefore the wavelength of luminescing photons  
          is {\it longer})
          for a {\it bigger} nanograin (see \S4 in Li 2004).
          }
    Observationally, Darbon et al.\ (1999)
    and Smith \& Witt (2002)
    showed that the ERE peak wavelength is indeed shifted toward 
    longer wavelengths with increasing UV radiation density.

Several materials have been proposed as ERE carriers: 
carbon nanoparticles (Seahra \& Duley 1999),
silicon nanoparticles 
(Witt et al.\ 1998, Ledoux et al.\ 1998; 
but see Li \& Draine 2002), 
nanodiamonds (Chang et al.\ 2006), 
and PAH clusters (Berne et al.\ 2008).
But none of them satisfies all the observational
requirements (see Li 2004).

\vspace{-3.0mm}
\section{Photoelectric Heating of the ISM Gas by Nanodust
            \label{sec:photoelectric}}
\vspace{-2.0mm}
In the Milky Way galaxy, approximately 60\% (by mass)
of the gas is in atomic hydrogen (HI) regions.
Observations of the 21\,cm line,\footnote{%
  The 21\,cm line originates in the hyperfine splitting 
  of the parallel and antiparallel spin states of
  the electron (relative to the spin of the proton)
  in the electronic ground state ($1s$) of atomic H. 
  }
show that the diffuse neutral atomic gas in the ISM is in two distinct phases
with temperatures $\simali$100$\K$ (``Cold Neutral Medium'', which 
accounts for, by mass, $\simali$40\% of the HI in HI regions)
and $\simali$6000$\K$ (``Warm Neutral Medium'', which 
accounts for $\simali$60\% of the HI in HI regions).
What heats the HI gas to temperatures of 
$\simali$100$\K$ or $\simali$6000$\K$?
One possible answer to this question provides indirect evidence 
for the existence of an appreciable quantity of
nanodust in the ISM.

\begin{figure}
\centering
\vspace{-2mm}
\includegraphics[width=9.8cm]{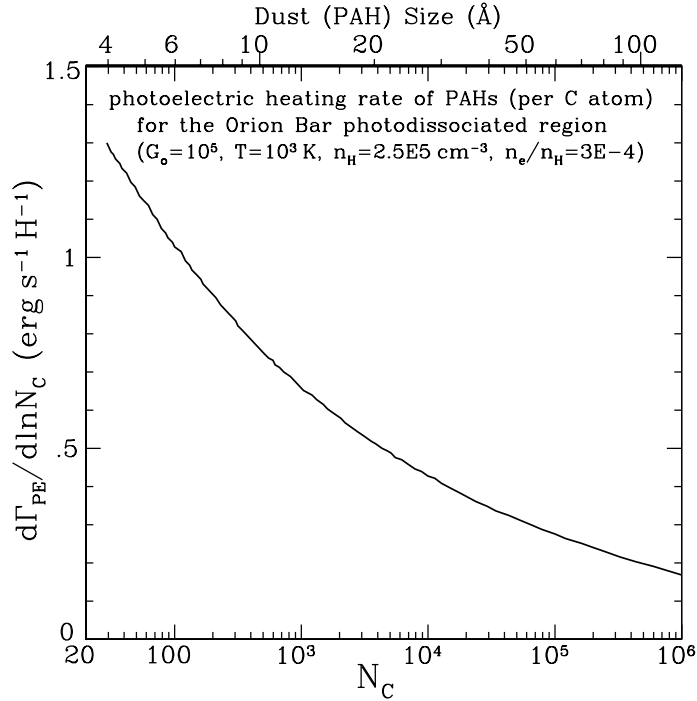}
\vspace{-3mm}
\caption{
         \label{fig:PE_Heating}
         The photoelectric heating rate ${\rm \Gamma}_{\rm PE}$ 
         of the interstellar gas in the Orion Bar 
         photodissociated region 
         as a function of PAH size (measured by $N_{\rm C}$, 
         the number of carbon atoms; the upper axis label gives 
         the graphite-equivalent spherical radius 
          $a$\,$\approx$\,$1.286\,N_{\rm C}^{1/3}\Angstrom$,
          or $N_{\rm C}$\,$\approx$\,$0.468\,\left(a/\Angstrom\right)^3$).
         The rates are presented in such a way that equal areas 
         under the curve correspond to equal contributions 
         to the heating. Typically, approximately half of 
         the heating originates from PAHs and PAH clusters 
         (with $N_{\rm C}$\,$<$\,$10^3$ or $a$\,$<$\,1.3\,nm).
         The other half is contributed by grains 
         of sizes 1.3\,nm\,$<$\,$a$\,$<$\,10\,nm).
         Larger grains do not contribute noticeably to the heating. 
         Data taken from Bakes \& Tielens (1994).
        }
\end{figure}

It has long been recognized that interstellar grains 
are an important energy source for heating the interstellar gas
through ejection of photoelectrons because 
(a) photons with energies below the ionization potential of H
    ($\simali$13.6$\eV$) do not couple directly to the gas; and 
(b) other heating sources such as cosmic rays, magnetic fields, 
    and turbulence are not important as a global heating source 
    for the diffuse ISM
    (e.g. the cosmic ray flux is too low 
     by a factor of $\simali$10 to account for 
     the interstellar gas heating; Watson 1972).

The photoelectric heating starts from the absorption
      of a UV photon by a dust grain, followed by
      ejection of an electron (``photoemission'').
The photoelectron diffuses toward the grain surface
and transfers (through inelastic collision) to the gas 
the excess (kinetic) energy left over after overcoming 
the work function 
(the binding energy of the electron to the grain) 
and the electrostatic potential 
(i.e. Coulomb attraction) of the grain (if it is charged).
 
In the diffuse ISM, nanodust (and in particular, 
      angstrom-sized PAH molecules) are much more efficient 
      in heating the gas than large grains 
      (see Tielens 2008)
since (a) the mean free path of an electron in a solid 
      is just $\simali$1$\nm$ and therefore photoelectrons
      created inside a large grain rarely reach the grain surface;
and (b) the total far-UV absorption is dominated by 
the nanodust component (see \S\ref{sec:IRemission}).
Theoretical studies have shown that grains smaller than $10\nm$
are responsible for $\gtsim 96\%$ of the total photoelectric 
heating of the gas, with half of this provided by
grains smaller than 1.5$\nm$ 
(Bakes \& Tielens 1994, Weingartner \& Draine 2001b).
Figure \ref{fig:PE_Heating} shows the calculated 
photoelectric heating rate as a function of grain size, 
illustrating that the photoelectric heating is 
dominated by the smallest grains (i.e. PAHs) 
present in the ISM.\footnote{%
  In the diffuse ISM, 
  PAHs and PAH clusters (or very small graphitic grains)
  dominate the nanodust population. 
  Nondetection of the 9.7$\mum$ silicate Si--O emission 
  feature in the diffuse ISM (see Figure \ref{fig:draine_irem})
  indicates that nano silicate grains are not abundant 
  (see Li \& Draine 2001b).
  } 
  
Observations confirm the dominant role 
of nanodust in the gas heating 
through the photoelectric effect. 
For instance, Habart et al.\ (2001) studied 
the major cooling lines, [CII]\,158$\mum$ and [OI]\,63$\mum$,
of L\,1721, an isolated cloud illuminated by a B2 star
in the $\rho$ Ophiuchi molecular complex.
Because of the energy balance between heating and
cooling, the [CII]\,158$\mum$ and [OI]\,63$\mum$
cooling lines (which dominate the gas cooling) 
reflect the heating input to the gas.
They found that the spatial distribution of the gas
cooling lines closely correlate with that of
the mid-IR emission attributed to 
nanodust (and PAHs). 

\begin{figure}
\centering
\includegraphics[width=10.8cm]{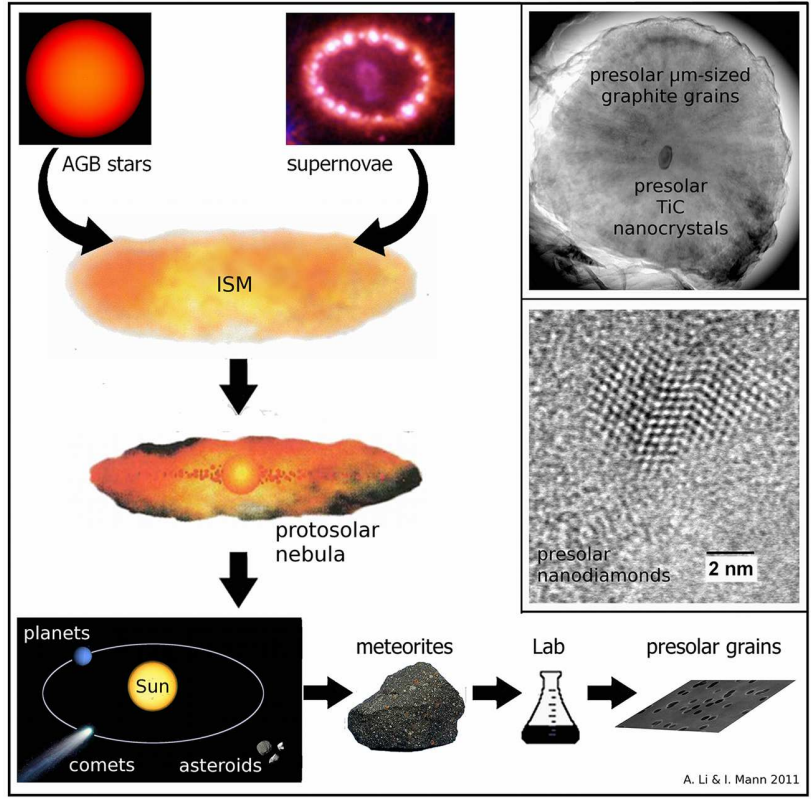}
\vspace{-2mm}
\caption{
         \label{fig:presolar_grains}
         A schematic illustration of the history of 
         presolar grains from their condensation in 
         stellar winds of AGB ({\it Assymptotic Giant Branch}) stars or 
         in supernova ejecta,
         to their injection into the ISM,
         and subsequent incorporation into 
         the dense molecular cloud 
         from which our solar system formed
         (i.e. protosolar nebula).
         These grains survived all the violent processes 
         occurring in the ISM (e.g. sputtering by shock waves)
         and in the early stages of solar system formation 
         and were incorporated into meteorite parent bodies.
        Experimental studies in terrestrial laboratories 
        allow to separate them from
        the meteorite or interplanetary dust material 
        in which they are embedded.         
	 Inserted are the TEM 
         ({\it Transmission Electron Microscopy}) 
         images of presolar nanodiamond grains
         (Banhart et al.\ 1998)
         and a presolar TiC nanocrystal within 
         a micrometer-sized presolar graphite spherule
         (Bernatowicz et al.\ 1991).
         }
\end{figure}

\vspace{-3.0mm}
\section{Direct Evidence: 
            Presolar Nanodust in Primitive Meteorites
            \label{sec:presolargrain}}
\vspace{-2.0mm}
Based on their isotopic anomalies, 
presolar grains (such as graphite, silicate, 
silicon carbide SiC, silicon nitride Si$_3$N$_4$,
and refractory oxides including corundum Al$_2$O$_3$, 
spinel MgAl$_2$0$_4$) that predate the solar system 
have been identified in primitive meteorites,
a class of meteorites that essentially remain 
chemically unaltered since their formation 
in the solar nebula (e.g. see Lodders 2005).
Presolar nanodiamonds of radii $a$\,$\simali$1\,nm 
(see Figure \ref{fig:presolar_grains})
were found to be rich in primitive carbonaceous meteorites, 
with an abundance as much as $\simali$0.1\% of the total mass 
in some primitive meteorites, more abundant than 
any other presolar grains by over two orders of magnitude
(Lewis et al.\ 1987).\footnote{%
   It is worth noting that, as mentioned earlier, 
   presolar grains are usually recognized by the isotope 
   ratios that are different from those for grains formed 
   in the solar system, but for small nanodiamonds, this 
   method does not work as accurately as for large grains 
   (since they are too small). There are debates that 
   a fraction of the meteoritic nanodiamonds may actually
   have formed in the early solar system (e.g. see Ott 2007).
   }
Presolar titanium carbide (TiC) nanocrystals 
were also seen in primitive meteorites 
(see Figure \ref{fig:presolar_grains}).
With a mean radius of $\simali$3.5\,nm,
they occur as nano-sized inclusions 
within micrometer-sized presolar graphitic
spherules (Bernatowicz et al.\ 1991). 

These presolar nanograins, 
after their condensation in stellar outflows
from carbon-rich evolved stars (e.g. TiC nano crystals)
or in ejecta from supernova explosion (e.g. nanodiamonds),
and prior to their incorporation into
the parent bodies of meteorites during 
the early stages of solar system formation,
they must have had a sojourn in the ISM   
out of which the solar system formed
(see Figure \ref{fig:presolar_grains} for illustration).

However, neither nanodiamonds nor TiC nanocrystals
could be representative of the bulk composition of 
interstellar nanograins and one observation is often 
explained in several different ways (Draine 2003).

\begin{figure}
\centering
\includegraphics[width=12.4cm]{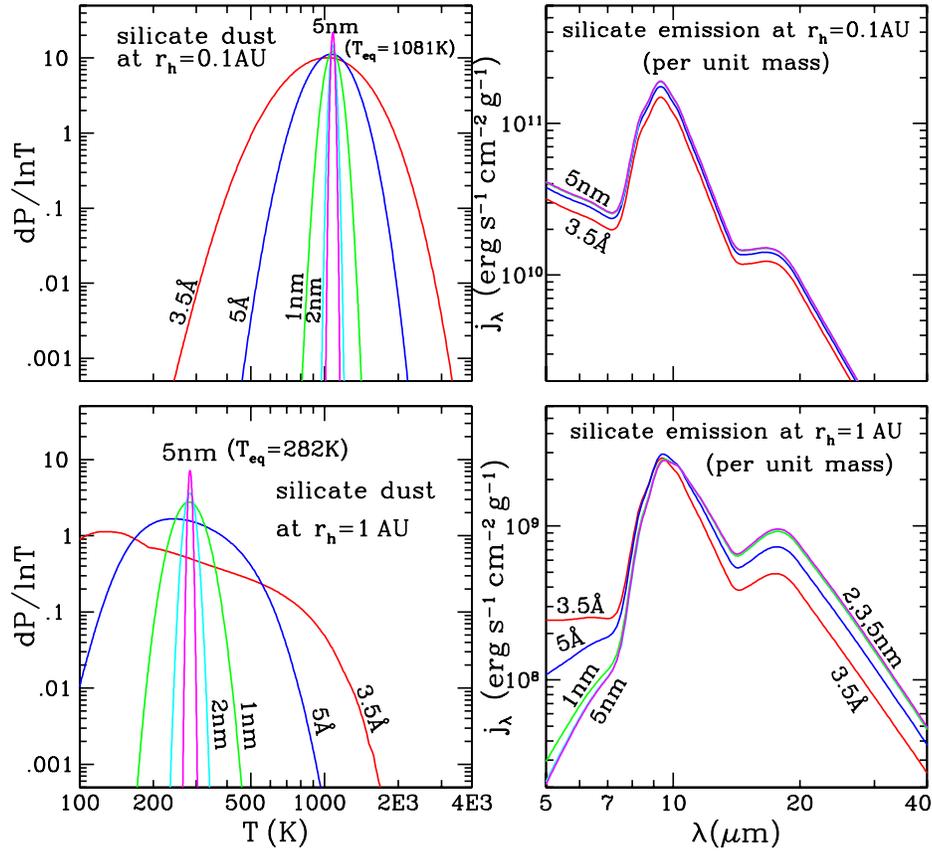}
\vspace{-8mm}
\caption{
         \label{fig:dPdTSun}
         Temperature probability distribution $dP/d\ln T$
         and emission for selected silicate grains
	 heated by the Sun at heliocentric distances
         of $r_{\rm h}$\,=\,0.1, 1$\AU$.
         The $dP/d\ln T$ distribution functions
         at $r_{\rm h}$\,=\,1$\AU$ are much broader
         than that for the same dust at $r_{\rm h}$\,=\,0.1$\AU$.  
         The grains with $a$\,$>$\,1\,nm peak at
         the same temperature (which is their equilibrium
         temperature), as expected from their Rayleigh 
         scattering nature (see \S\ref{sec:SolarSystem}).
         }
\end{figure}
      

\section{Summary and Comparison to Nanodust in the Solar System
            \label{sec:SolarSystem}}

Based on the discussions in the previous sections 
we can summarize the observational
evidence for the presence of nanodust in the ISM:
(i) the far-UV extinction at $\lambda^{-1}$\,$>$\,6$\mum^{-1}$
    caused by the absorption of nanodust
    (\S\ref{sec:extinct}),
(ii) the $\simali$2--60$\mum$ near- and mid-IR 
    spectral and continuum emission 
    from stochastically heated nanograins
    through vibrational relaxation (\S\ref{sec:IRemission}),
(iii) the $\simali$10--100\,GHz ``anomalous'' Galactic 
     foreground microwave emission from rotationally 
     excited nanograins through electric dipole
     radiation (\S\ref{sec:microwave}), and
(iv) the $\simali$5400--9500$\Angstrom$ broad,
      featureless ERE band from electronically excited 
      nanograins through photoluminescence (\S\ref{sec:ERE}).
  The presence of an appreciable amount of nanograins 
in the ISM is also indirectly inferred from 
the photoelectric heating of interstellar gas
by nanodust (i.e. PAHs; \S\ref{sec:photoelectric}) and finally
presolar grains are found in interplanetary dust 
and meteorites of the solar system.

Finally, it is worth noting that the nanodust
population (particularly PAHs) may be responsible 
for the lower gas-phase deuterium abundance 
of D/H\,$\approx$\,7--22\,ppm in the Galactic ISM 
compared to the primordial value of D/H\,$\approx$\,26\,ppm, 
through depleting the ``missing'' D onto PAHs (Draine 2006).

Based on the preceding discussion the physical processes that
allow for (direct or indirect) observation of nanodust are
UV light-scattering,
stochastic heating, 
electric dipole
     radiation of rotating nanodust,
photoluminescence, 
      and
finally photoelectric heating of the surrounding gas.      

We now discuss the possible occurrence of these processes
in the solar system.      
It is worth noting at the beginning, that the nanodust 
in the solar system is not identical to 
the nanodust in the ISM. 
The grains with $a$\,$<$\,5\,nm in the local interstellar cloud 
in the vicinity of the Sun do not enter the solar system, 
as they are deflected by the magnetic field that builds
up in the outer solar system due to interaction of 
the local ionized ISM gas with the solar wind (Mann 2010).
The nanodust in the solar system is produced locally
from the interplanetary dust cloud or from solar system objects.


\begin{itemize}
\vspace{-1.0mm}
\item {\bf Light-scattering in far UV:}
      Scattering from interplanetary dust particles 
      generated the Zodiacal light,
      which is the predominant diffuse brightness 
      to wavelengths as short as $\lambda$\,$\simali$0.3$\mum$. 
      The brightness at shorter wavelength is dominated by
      the emission of unresolved stars \citep{Leinert1998}. 
      Dust emission in the X-ray  was discussed after 
      the ROSAT survey established comets as a class 
      of X-ray sources. The discussed dust-related X-ray 
      signals are X-ray fluorescence and scattering by 
      nano dust and X-ray emission caused by high-velocity 
      impacts of nano dust.
      The X-ray flux from nanodust in the solar system 
      is estimated by Kharchenko and Lewkow in this book. 

\vspace{2.0mm}
\item {\bf Stochastic heating:}
     For the nanograins in the inner solar system (Mann 2007),
     it is more likely for them to attain an equilibrium temperature
     (compared to the same dust in the diffuse ISM).
     This is because the solar system nanodust is exposed to 
     a far more intense radiation field: 
     at a heliocentric distance of $r_{\rm h}$,
     the 912$\Angstrom$--1$\mum$ solar radiation intensity is
     $\approx$\,$7.6\times 10^7\,\left(r_{\rm h}/{\rm AU}\right)^{-2}$
     times that of the local interstellar radiation field, 
     where $r_{\rm h}$ is 
     the heliocentric distance of the dust from the Sun.   
     
\vspace{2mm}
In Figure \ref{fig:dPdTSun} we present the temperature
probability distribution functions of amorphous silicate
dust using the dielectric functions of Draine \& Lee (1984)
for selected sizes
($a$\,=\,3.5\AA, 5\AA, 1$\nm$, 2$\nm$, 5$\nm$) illuminated
     by the Sun at 
     $r_{\rm h}$\,=\,0.1,\,1$\AU$.   
To facilitate comparison, we plot $dP/d\ln T$
in the same $T$ and $dP/d\ln T$ ranges
for $r_{\rm h}$\,=\,0.1$\AU$
and $r_{\rm h}$\,=\,1$\AU$.
The emission is also illustrated in the same
$\lambda$ and $j_{\lambda}$ ranges 
(except the latter differs by a factor of $r_{\rm h}^{-2}$).  
We see that silicate dust with $a$\,$\simgt$\,2\,nm
attains an equilibrium temperature of 
$T_{\rm eq}$\,$\approx$\,282$\K$ at $r_{\rm h}$\,=\,1$\AU$.  
At $r_{\rm h}$\,=\,0.1$\AU$, the $dP/d\ln T$ 
     distribution functions for grains as small as
     $a$\,=\,1\,nm is already like a delta function,
     peaking at  $T_{\rm eq}$\,$\approx$\,1081$\K$.
     At $r_{\rm h}$\,=\,1$\AU$, the silicate emission
     spectra for $a$\,=\,2,\,3,\,5\,nm are almost
     identical. This is because in the entire UV to far-IR 
     wavelength range, these nanograins are in the Rayleigh
     regime and their $Q_{\rm abs}/a$ values are independent
     of size, therefore they obtain an almost identical equilibrium
     temperature. At $r_{\rm h}$\,=\,0.1$\AU$, this even applies
     to smaller grains (e.g. $a$\,=\,1\,nm).    
To summarize, in the solar system at $r_{\rm h}$\,$<$\,1\,AU,
the stochastic heating effect is small for dust larger than 
$\simali$1\,nm in radius; for dust smaller than $\simali$5$\Angstrom$,
it may not survive since the stochastic heating would lead to 
temperatures exceeding $\simali$2000\,K.
This is roughly the sublimation temperature of silicate dust, though
we point out that the nanodust can have lower sublimation temperature
than the bulk material (see Kimura, this book).

%
\vspace{2.0mm}
\item {\bf Electric dipole radiation of rotating dust:}
      We are not aware of any studies of the rotational dynamics
      of nanodust in the solar system. We do not expect to see
      strong microwave emission from the dust in the solar system.
      Although the ions in the solar system may deliver more angular
      momentum to a grain than in the diffuse ISM
      (because of their large abundance and large kinetic energy),
      the dust will not be driven to rotate 
      as fast as in the diffuse ISM, due to 
      (1) the large grain size of the solar system nanodust population,
      (2) the strong rotational damping caused by photon emission
         in the solar system, and 
      (3) the small number densities of nanodust in the directions 
          facing away from the Sun (see other chapters of this book). 
  The nanodust in the solar system seems to be in
  the nm size range (see other chapters of this book), 
  while the microwave emission in the ISM arises
  predominantly from angstrom-sized PAHs.
  Note that the angular velocity $\omega$\,$\propto$\,$1/I$,
  while $I$\,$\propto$\,$a^5$.

\vspace{2.0mm}
\item {\bf Photoluminescence:}
As far as observations in the solar system 
are concerned a coronal emission that appeared similar 
to the ERE was speculated to result from silicon nano crystals 
near the sun (Habbal et al.\ 2003), 
but this was challenged on the basis of 
the dust composition and emission properties (Mann \& Murad 2005). 
As opposed to the ISM the nanodust in the solar system most
likely is heterogeneous in composition and covers a broad size interval. 
This broadens wavelength at which the photoluminescence is observed,
which makes its detection near the Sun less likely.  
At this point we are not aware of an observational 
result in the optical or IR range confirming 
the existence of nano dust in the solar system 
and the detection of nano dust in the interplanetary medium 
in a similar way like the ERE is unlikely 
(see Mann \& Czechowski 2011 in this book).

\vspace{2.0mm}
\item {\bf Photoelectric heating:} 
For the solar system nanodust, photoelectric heating 
does not occur because the nanodust is embedded 
in the high temperature solar wind electrons. 
Instead, the presence of nanodust in the solar wind 
would rather lead to cooling, because the solar wind 
ions charge-exchange and are decelerated when passing a nanograin. 
\end{itemize}      

\vspace{-5.0mm}   
\section{Conclusion\label{sec:Conclusion}}
\vspace{-2.0mm}
Observational data, particularly the near- and mid-IR
emission data, allow us to constrain the composition, 
size distribution and quantity   
of the nanodust population in the ISM:
(i) PAHs and nano-sized graphitic grains
    are the dominant nanodust species in the ISM;
    locking up $\simali$15\% of the total interstellar carbon,
    they are responsible for the IR emission 
    at $\lambda$\,$<$\,60$\mum$ 
    (including the ``UIR'' bands)
    and the microwave emission as well;
(ii) nano silicate grains are not important:
     they {\it at most} account for $\simali$5\%
     of the total interstellar silicon
     as indicated by the nondetection of
     the 9.7$\mum$ silicate Si--O emission band
     in the Galactic diffuse ISM
     (Li \& Draine 2001b).
Presolar nanograins
(i.e. nanodiamonds and TiC nanocrystals) that are
identified in primitive meteorites and interplanetary dust
are not an abundant population of the ISM nanodust, but
provide the most direct evidence 
for interstellar nanodust. They were present in the
local ISM at the time of the formation of the solar system.
The physical processes involving nanodust in the ISM
are less important for the nanodust in the solar system.
Though considering the stochastic heating process suggests
that the size of the silicate nanodust in the inner solar system
is constrained to $a$\,$\gtsim$0.5$\nm$.

%
%


%

\begin{figure}
\centering
\hspace{-2mm}
\includegraphics[height=10.6cm, angle=90]{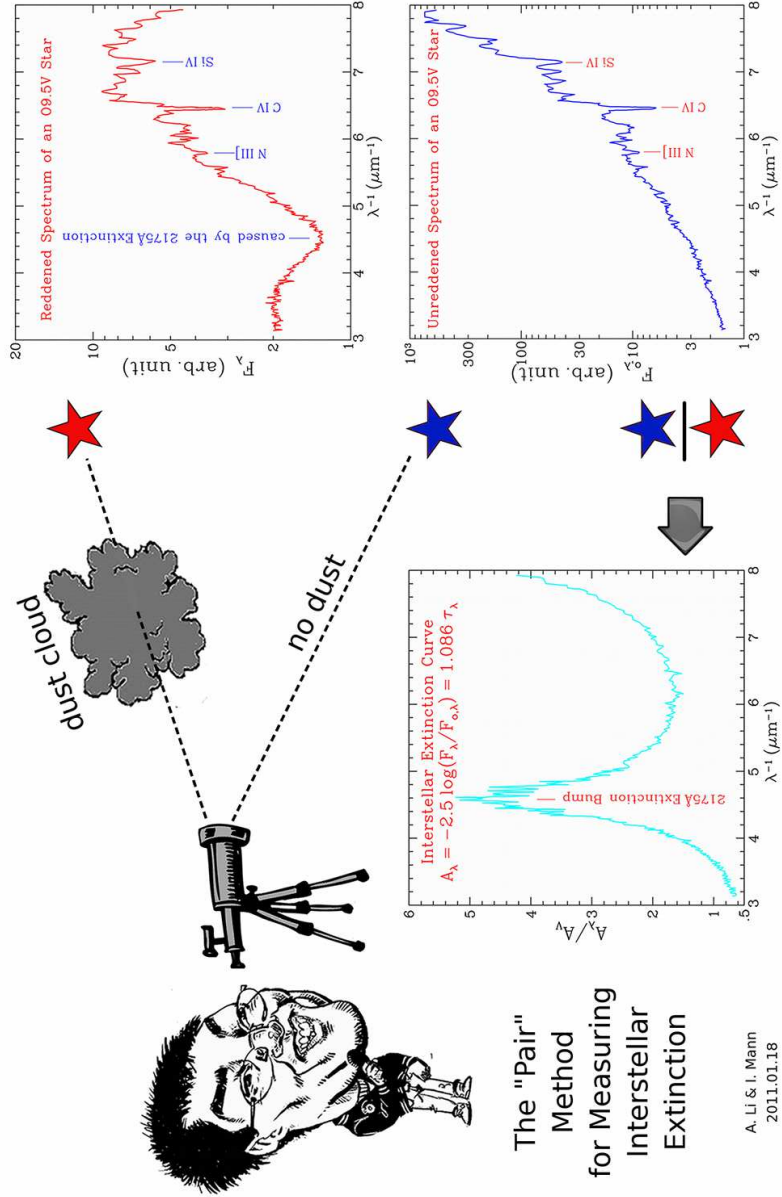}
\vspace{-2mm}
\caption{
         \label{fig:extcurv_cartoon}   
         A conceptual illustration of the ``pair'' method.
                  }
\end{figure}

\section*{Appendix: Interstellar Extinction}
For the ISM in the solar neighborhood 
(i.e. a few kiloparsecs from the Sun and within $\simali$100\,pc
of the galactic plane), the mean visual extinction 
has long been determined quite accurately.
The interstellar extinction curve is most commonly derived
utilizing the ``pair'' method. 
As illustrated in Figure \ref{fig:extcurv_cartoon},
this technique involves photometric or spectrophotometric 
observations of two stars of identical spectral types,
with one star located behind a dust cloud and 
another star, (in ideal case) unaffected by interstellar dust,
so that there is no obscuration between the observer and the star.
Let $F_\lambda$ be the observed flux from the reddened star,
and $F_{\rm o,\,\lambda}$ be the flux from the unreddened star.
If both stars are located at an identical distance, 
the extinction $A_\lambda$ -- measured in ``magnitudes'' 
-- is 
\begin{equation} 
A_\lambda \equiv 2.5 \log_{10}
\left[F_{\rm o,\,\lambda}/F_\lambda\right]
\approx 1.086 \tau_\lambda
\end{equation} 
%
%
where $\tau_\lambda$ is the optical depth.  
As it is often not possible to find reddened/unreddened 
star pairs of identical spectral types which are also
located at identical distances, one often measures the 
color excess 
\begin{equation} 
E(\lambda-V)\equiv A_\lambda-A_V
= 2.5 \log_{10}
\left[\frac{F_{\rm o,\,\lambda}/F_{\rm o,\,V}}
{F_\lambda/F_V}\right]
\end{equation}
from normalized stellar fluxes,
with the $V$ band being the usual choice 
for the normalization purpose.
The total-to-selective extinction ratio

\begin{equation} 
R_V \equiv A_V/E(B-V) 
\end{equation}
suggested by (Cardelli et al.\ 1989) is frequently used
to characterize the galactic extinction curves. 

For particles  in the Rayleigh regime
(i.e. $2\pi a/\lambda$$\ll$1) their extinction cross sections 
per unit volume, $C_{\rm ext}(a,\lambda)/V$,
are independent of size.
Therefore, the observational quantity $A_\lambda/\NH$ 
($\magni\cm^{-2}\H^{-1}$) -- the extinction per unit H column
-- only constrains $V_{\rm tot}/\nH$, the total dust volume per H nuclei 
of this grain component:

\begin{equation} 
A_\lambda/\NH = 1.086\,\int C_{\rm ext}(a,\lambda)\, 
n_{\rm H}^{-1} \left(dn/da\right)da
= 1.086\,\left(V_{\rm tot}/\nH\right) \left(C_{\rm ext}/V\right),
\end{equation}
where $dn$ is the number of grains in the size interval $[a,\,a+da]$,
and $\NH$ ($n_{\rm H}$) is the hydrogen column (volume) density. 
This explains why the MRN silicate-graphite model 
with a lower size cut-off of 
$a_{\rm min}$\,=\,5$\nm$ (Mathis et al.\ 1977),
that was frequently used before the presence of nanodust was confirmed,
could also closely reproduce 
the observed extinction curve. 
  The MRN model fitted the extinction curve using a mixture
  of silicate and graphite grains with a simple power-law
  size distribution: $dn/da \propto a^{-3.5}$ 
  for 5$\nm$$\simlt$\,$a$\,$\simlt$0.25$\mum$.

The 2175$\Angstrom$ bump is thought to be predominantly 
due to absorption,  as indicated by the broad minimum 
near 2175$\Angstrom$ of the interstellar albedo (Whittet 2003).
The interstellar albedo is defined as the ratio of 
scattering to extinction. The interstellar extinction is 
the combination of scattering and absorption.
For grains in the Rayleigh limit, 
the scattering is negligible in comparison 
with the absorption (see Li 2008).
This suggests that its carrier is sufficiently small 
to be in the Rayleigh limit,
with a size $a$\,$\ll$\,$\lambda/2\pi$\,$\approx$\,35$\nm$.

Finally, we should note that a smooth extension of 
the MRN $dn/da$\,$\propto$\,$a^{-3.5}$ size distribution 
down to $a$\,=\,3$\Angstrom$ is not sufficient to account 
for the observed 12$\mum$ and 25$\mum$ emission 
(see Draine \& Anderson 1985, Weiland et al.\ 1986) and that
  an extra population of nano-sized dust is required
  (e.g. see D\'esert et al.\ 1990, Dwek et al.\ 1997, Li \& Draine 2001a).
In the recent Weingartner \& Draine (2001a) model, 
for instance, the grain size distribution 
extends from a few angstroms to a few micrometers,
with $\simali$6\% of the total dust mass in 
grains smaller than 2$\nm$.

\section*{Appendix: Stochastic Heating}
%
%
Typically large dust particles in the interstellar medium 
reach equilibrium temperature for which the rate of radiative cooling 
equals the time-averaged rate of energy absorption.
If the emissivity of the dust material is roughly constant with
wavelength (which is the case for a sufficiently large dust particle)
then the spectral slope of the thermal emission brightness is 
roughly that of a black body (Plankc curve) with the location
of maximum emission being determined by the temperature. 
For interpreting astronomical observations a dust temperature
is often assumed to be the equilibrium temperature and one denotes
as colour temperature of an object the temperature of a blackbody with
peak emission at the same wavelength as the observed 
brightness.
The required dust temperature 
to generate peak emission in the mid-IR
is, for instance $\simali$300$\K$ for the 12$\mum$ emission 
and $\simali$150$\K$ for the 25$\mum$ emission.
When discussing nanodust, researchers often use the Debye 
temperature $\Theta$ which is a parameter 
(with a dimension of kelvin) that characterizes 
the low-temperature heat capacity $U$ of a solid.

Nanograins are small enough that their time-averaged 
internal energy is smaller than or comparable to 
the energy of the starlight photons that heat the grains. 
Stochastic absorption of photons 
therefore results in transient ``temperature spikes'', 
during which much of the energy deposited 
by the starlight photon is reradiated 
in the near- or mid-IR.

The ``temperature spike'' -- the maximum temperature 
to which a nanograin can reach upon an absorption of 
a photon is sensitive to 
its heat capacity (which is $\propto$\,$a^3$). 
When illuminated by a radiation field,
the observed intensity $I_\lambda$ 
from the transiently heated nanograins
is 
\begin{equation} 
I_\lambda = \NH \int C_{\rm abs}(a,\lambda)\,n_{\rm H}^{-1} 
\left(dn/da\right)da \int B_\lambda(T)\left(dP/dT\right)_a dT
\end{equation} 
where $C_{\rm abs}(a,\lambda)$ is the absorption cross section 
for a grain of radius $a$ at wavelength $\lambda$,
$B_\lambda(T)$ is the Planck function at temperature $T$, and
$\left(dP/dT\right)_a$ is the dust temperature distribution function
which is sensitive to grain size $a$ (see Li 2004).
We see that although in the IR wavelength range 
$C_{\rm abs}(a,\lambda)/V$ is independent of grain size $a$,
$\left(dP/dT\right)_a$ is a sensitive function of $a$ 
and therefore $I_\lambda$ allows us to constrain 
the size distribution of the nanodust component 
through $\left(dP/dT\right)_a$. 

\begin{figure}
\centering
\vspace{-1mm}
\includegraphics[angle=270,width=11.6cm]{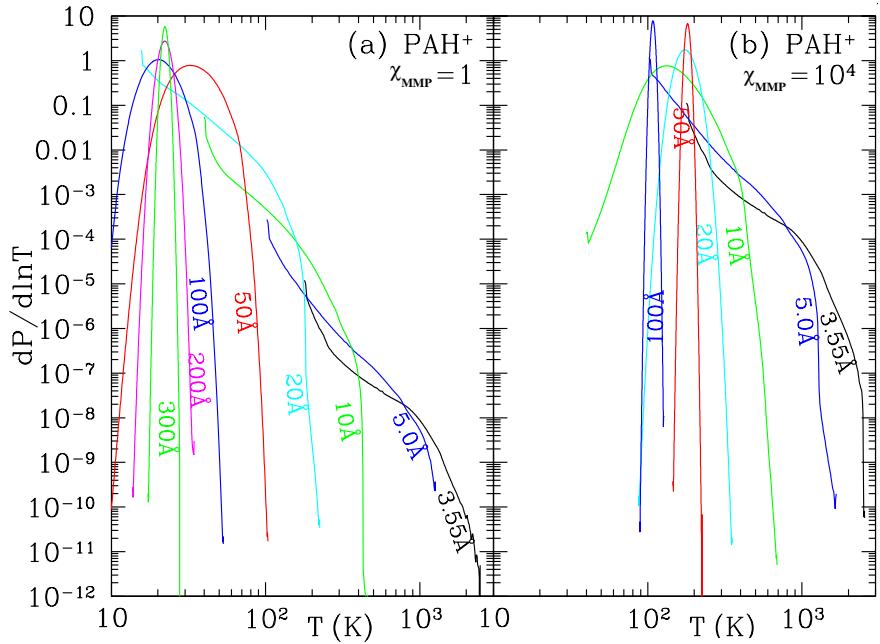}
\vspace{-2mm}
\caption{
         \label{fig:dPdlnT}
         Temperature probability distribution $dP/d\ln T$ 
         for selected PAH ions
	 heated by starlight with $\chiMMP$\,=\,1 
         and $\chiMMP$\,=\,10$^4$
         (in unit of the MMP local interstellar radiation field).
         It is interesting to note that
         in the case of $\chiMMP$\,=\,1,
         the $a$\,=\,20$\nm$ grain
         and the $a$\,=\,30$\nm$ grain
         (even the $a$\,=\,10$\nm$ grain as well)
         peak at the same temperature
         (which is their equilibrium temperature).
         This is expected from their Rayleigh scattering nature
         (see the caption of Figure \ref{fig:Tss}).
         However, in the case of $\chiMMP$\,=\,10$^4$,
         the $a$\,=\,5$\nm$ grain
         and the $a$\,=\,10$\nm$ grain
         do not peak at the same temperature,
         although they both attain equilibrium temperatures.
         This is because they were ``designed'' to have different 
         optical properties:
         PAHs with $a$\,$\gtsim$\,10$\nm$ have graphitic
         properties, while those with $a$\,$\ltsim$\,5$\nm$ 
         have PAH properties (see Li \& Draine 2001a). 
         Taken from Draine \& Li (2007).    
         }
\end{figure}

Figure \ref{fig:dPdlnT} shows the temperature 
probability distributions 
$dP/d\ln T$ 
for PAH ions of selected sizes 
(at $a$\,$>$\,5$\nm$ their optical properties 
approach that of graphite; see Li \& Draine 2001a).
The distributions are shown for $\chiMMP$\,=\,1 and $\chiMMP$\,=\,$10^4$,
where $\chiMMP$ is the starlight intensity 
in units of the interstellar radiation field
given by Mathis, Mezger, \& Panagia (1983) (MMP).
   We see in Figure \ref{fig:dPdlnT}a that small grains 
   undergo extreme temperature excursions 
   (e.g. the $a$\,=\,3.55$\Angstrom$ PAH 
   occasionally reaches $T$\,$>$\,2000$\K$), 
   whereas larger grains (e.g. $a$\,=\,30$\nm$) 
   have temperature distribution functions that are 
   very strongly-peaked and like a delta-function,
   corresponding to only small excursions around 
   an equilibrium temperature. 
It is apparent that when 
    the rate of photon absorption increases,
    the ``equilibrium'' temperature approximation 
    becomes valid for smaller grains;  
    e.g., for $\chiMMP$\,=\,$10^4$ one could 
    approximate a $a$\,=\,5$\nm$ grain as having 
    an equilibrium temperature $T_{\rm eq}$\,$\approx$\,150$\K$
    whereas for $\chiMMP$\,=\,1 the temperature excursions 
    are very important for this grain
    (see Draine \& Li 2007).
    This is also the reason for stochastic heating not being 
    important in the inner solar system.

Whether a grain will undergo stochastic heating depends on
    (i) the grain size,
    (ii) the optical properties of the dust,
    (ii) the thermal properties (e.g. Debye temperature) of the dust,  
    (iii) the starlight intensity, and
    (iv) the hardness of the starlight, which measures the relative
    amount of short-wavelength (``hard'') photons compared to 
    long-wavelength (``soft'') photons.
     For a {\it smaller} grain with a {\it smaller}
     UV/visible absorptivity 
     and a {\it larger} Debye temperature ${\rm \Theta}$,
     exposed to starlight of a {\it lower} intensity
     and a {\it harder} spectrum, it is {\it more} likely
     for this grain to be stochastically heated by single photons.
     This is because (i) the specific heat of a grain 
     (at a given temperature) is proportional to $a^3/{\rm \Theta}^3$,
     a single photon (of a given energy) would therefore 
     result in a {\it more} prominent temperature spike for
     a {\it smaller} grain with a {\it larger} ${\rm \Theta}$;
     (ii) the photon absorption rate is proportional to
     the starlight intensity and the absorptivity of the dust
     in the UV/visible wavelength range, a {\it smaller} grain
     with a {\it smaller} UV/visible absorptivity when exposed
     to a more {\it dilute} radiation field will 
     have a {\it smaller} photon absorption rate
     and will therefore more likely undergo stochastic heating;
     (iii) a {\it more} energetic photon would cause a grain
     to gain a {\it larger} temperature raise, grains are therefore
     {\it more} likely to experience transient heating when
     exposed to a {\it harder} radiation field.

\vspace{3mm}
{\noindent {\bf Acknowledgements}~ We thank M. K\"ohler
for her great help in preparing for 
Figures \ref{fig:presolar_grains},\,\ref{fig:extcurv_cartoon}.
We thank B.T. Draine, J. Gao, B.W. Jiang,
S. Kwok and N. Meyer-Vernet for their 
helpful comments and suggestions.
We also thank A.N. Witt and U.P. Vijh for providing us
the ERE data of NGC\,7023,
F. Banhart and T.J. Bernatowicz respectively 
for providing us the TEM images of 
presolar nanodiamonds and presolar graphite
with TiC nanocrystals embedded. A.L. is supported in part
by a NSF grant (AST-1109039) and a NASA Herschel Theory program.} 



\printindex


\end{document}